\documentclass[12pt]{article}

\RequirePackage{graphicx}

\usepackage{amsmath}

\usepackage{amsfonts}

\RequirePackage{amssymb}

\RequirePackage{hyperref}

\RequirePackage[latin1]{inputenc}

\RequirePackage{epsfig}

\RequirePackage{lineno}

\RequirePackage{mathrsfs}

\RequirePackage[T1]{fontenc}

\RequirePackage{mathptmx}      
\RequirePackage{latexsym}
\RequirePackage[numbers,sort&compress]{natbib}
\newcommand{\h}{\hspace{.5cm}}
\date{}
\begin{document}

\title{Dirac $\delta$-function potential in quasiposition representation of a minimal-length scenario
}


\author{M. F. Gusson, A. Oakes O. Gon\c{c}alves, R. O. Francisco\\ R. G. Furtado, J. C. Fabris\footnote{julio.fabris@cosmo-ufes.org} \hspace{0.01mm} and J. A. Nogueira\thanks{jose.nogueira@ufes.br}\\
       [0.1cm]\small\emph{Departamento de F\'isica, Centro de Ci\^encias Exatas}\\[-0.1cm]\small\emph{Universidade Federal do Esp\'irito Santo -- UFES}\\[-0.1cm]\small\emph{29075-910 -- Vit\'oria -- ES -- Brasil}
}

\maketitle

\begin{abstract}

A minimal-length scenario can be considered as an effective description of quantum gravity effects. In quantum mechanics the introduction of a minimal length can be accomplished through a generalization of Heisenberg's uncertainty principle. In this scenario, state eigenvectors of the position operator are no longer physical states and the representation in momentum space or a representation in a quasiposition space must be used. In this work, we solve the Schroedinger equation with Dirac $\delta$-function potential in quasiposition space. We calculate the bound state energy and the coefficients of reflection and transmission for scattering states. We show that leading corrections are of order of the minimal length $({\sl O}(\sqrt{\beta}))$ and the coefficients of reflection and transmission are no longer the same for the Dirac delta well and barrier as in ordinary quantum mechanics. Furthermore, assuming that the equivalence of the 1s state energy of the hydrogen atom and the bound state energy of the Dirac $\delta$-function potential in 1-dim is kept in a minimal-length scenario, we also find that the leading correction term for the ground state energy of the hydrogen atom is of order of the minimal length and $\Delta x_{min} \le 10^{-25}$ m.

\end{abstract}

\pagenumbering{arabic}


\section{Introduction}
\label{introd}
\h Gravity quantization has become a huge challenge to theoretical physicists. Despite enormous efforts employed, so far, it was not possible to obtain a theory which can be considered suitable and not even a consensus approach. Nevertheless, most of the candidate theories to gravity quantization seem to have one common point: the prediction of the existence of a minimal length, that is, a limit for the precision of a length measurement.

Although the first proposals for the existence of a minimal length were done by the beginning of 1930s \cite{krag1,krag2,Heisenberg}, they were not connected with the quantum gravity, but instead of this with a nature cut-off that would remedy cumbersome divergences arising from quantization of systems with an infinite number of degrees of freedom. The relevant role that gravity plays in trying to probe a smaller and smaller region of the space-time was recognized by M. Bronstein \cite{Bronstein} already in 1936, however his works did not attract a lot of attention. It was only in 1964 that C. A. Mead \cite{Mead1,Mead2} once again proposed a possible connection between gravitation and minimal length. Hence, we can assume that gravity may lead to an effective cut-off in the ultraviolet. Furthermore, if we are convinced that gravitational effects are considered when a minimal length is introduced then a minimal-length scenario could be thought of as an effective description of quantum gravity effects \cite{Hossenfelder1}.

As far as we know, the introduction of a minimal-length scenario can be carried out through three different way \cite{Hossenfelder1,tawfik1,tawfik2}: a generalization of the Heisenberg's uncertainty principle (GUP), a deformation of the special relativity\footnote{It is named doubly special relativity because of the existence of two universal constants: light speed and minimal length.} (DSR) and a modification of the dispersion relation (MDR). 

Various problems connected with the minimal length have been studied in the context of the non-relativistic quantum mechanics. Among them are the harmonic oscillator \cite{Kempf:1994su,Kempf2:1997,chang1,dadic,Hassanabadi}, the hydrogen atom \cite{brau,yao,benczik,Stetsko,nouicer,fityo,bouaziz},  step and barrier potentials \cite{das1,das2,sprenger}, finite and infinite square wells \cite{nozari,blado}, as well as others.
In the relativistic context, the Dirac equation has been studied in \cite{nozari1,tkachuk,nouicer1,chargui,hassanabadi1,panella,panella1}. The Casimir Effect has also been studied in a minimal-length scenario in \cite{nouicer2,harbach,panella2,panella3,dorsch}.

An interesting problem in quantum mechanics is the Dirac $\delta$-function potential. In general, the Dirac $\delta$-function potential is used as a pedagogical exercise. Nevertheless, it has also been used to model physical quantum systems \cite{robinett}. Maybe because the attractive Dirac $\delta$-function potential is one of the simplest quantum system which displays both a bound state and a set of continuous states, it has been used to model atomic and molecular systems \cite{kronig,frost,frost1,frost2,kuhn,kuhn1}. In addition, the short-range interactions in condensed matter with a large scattering length can actually be modeled as a Dirac $\delta$-function potential \cite{tan,tan1,tan2,braaten,zhang}. In quantum field theory, in order to treat the Casimir effect more realistically, the boundary conditions are replaced by the interaction potential $\frac{1}{2} \sigma(x) \phi^{2}(x)$, whrere $\sigma(x)$ represents the field of the material of the borders (background field). Hence, for sharply localized borders  the background field can be modeled by a Dirac $\delta$-function \cite{jaffe,kimball}.

The Dirac $\delta$-function potential by its very nature is challenging problem in a minimal-length scenario.

N. Ferkous \cite{Ferkous} and M. I. Samar \& V. M. Tkachuck \cite{samar} have independently calculated the bound state energy in ``momentum space''. In both papers, the authors have found a correction for the expression of energy in $\sqrt{\beta}$ and $\beta$ ($\sqrt{\beta} \sim \Delta x_{min}$), but with different coefficients, therefore disagreeing outcomes.  M. I. Samar and V. M. Tkachuck claim that is because, whereas they consider $p$ belongs to $\left(-\frac{\pi}{2\sqrt{\beta}}, \frac{\pi}{2\sqrt{\beta}} \right)$, Ferkous consider $p$ belongs to $\left(-\infty, \infty \right)$. In this work, we propose to solve the problem of a non-relativistic particle of mass $m$ in the presence of Dirac $\delta$-function potential in quasiposition space. Since the quasiposition space representation is used we can consider the cases of bound states and scattering states as well. We find the same expression for the energy of the bound state obtained by Ferkous.

In addition, assuming that the equality between the 1s state energy of the hydrogen atom and the bound state energy of the Dirac $\delta$-function potential in 1-dim when the coefficient of the $\delta$-potential is replaced by the fine structure constant \cite{frost} is kept in a minimal-length scenario, we find that the leading correction for the ground state energy of hydrogen atom is of order of the minimal length $({\sl O}(\sqrt{\beta}))$, differently from commonly found in the literature using perturbative methods \cite{brau,yao,benczik,Stetsko,antonacci}, but in according to the results obtained by T. V. Fityo, I. O. Vakarchuk and V. M. Tkachuk \cite{fityo} and D. Bouaziz \& N. Ferkous \cite{bouaziz} using a non-perturbative approach.

The rest of this paper is organized as follows. In section \ref{MLS} we show how to introduce a minimal-length scenario and find the time-independent Schroedinger equation in quasiposition space representation. In section \ref{Ddfp} we solve the modified Schroedinger equation and find the bound state energy and the coefficients of reflection and transmission for the scattering states. We present our conclusions in section \ref{Concl}.


\section{Minimal-length scenario}
\label{MLS}
\h In quantum theory, a minimal-length scenario can be accomplished by imposing a non-zero minimal uncertainty in the measurement of position which leads to generalized uncertainty principle (GUP). Since
\begin{equation}
\Delta x \Delta p \geq \frac{|\langle [\hat{x},\hat{p}] \rangle |}{2},
\end{equation}
a generalization of the uncertainty principle corresponds to a modification in the algebra of the operators. There are different suggestions of modification of the commutation relation between the position and momentum operators which implement a minimal-length scenario. We concern with the most usual of them, proposed by Kempf~\cite{Kempf:1994su,Kempf2:1997}, which in a 1-dimensional space is given by
\begin{equation}
	\label{rc1kempf}
	[\hat{x},\hat{p}] := i\hbar \left(1 + \beta\hat{p}^2 \right),
\end{equation}
where $\beta$ is a parameter related to the minimal length. The commutation relation (\ref{rc1kempf}) corresponds to the GUP
\begin{equation}
	\label{string_uncertainty}
	\Delta x \Delta p \geq \frac{\hbar}{2}  \left[ 1 +  \beta (\Delta p)^2 + \beta \langle \hat{p}  \rangle^{2}  \right] ,
\end{equation}
which implies the existence of a non-zero minimal uncertainty in the position $\Delta x_{min} = \hbar \sqrt{\beta}$.

Unfortunately, in this scenario the eigenstates of the position operator are not physical sates\footnote{That is because the uncertainty $\Delta A$ of an operator $\hat{A}$ in any of its state eigenvectors $| \psi_{A} \rangle $ must be zero, which is not the case for the position operator, since $\Delta x_{min} > 0$.} and, consequently, the representation in position space can no longer be used, that is, an arbitrary state vector $| \psi \rangle $ can not be expanding in the basis of state eigenvectors of the position operator $\{| x \rangle \}$. Hence the obvious way ahead is to make use of the representation in momentum space:
\begin{equation}
\label{p-repx}
\langle p | \hat{x} | \psi \rangle = i \hbar \left( 1 + \beta p^{2} \right)\frac{\partial \tilde{\psi} (p)}{\partial p},
\end{equation}
\begin{equation}
\label{p-repp}
\langle p | \hat{p} | \psi \rangle = p \tilde{\psi} (p).
\end{equation}

However, the representation in momentum space is not suitable in some cases, such as, for example, when the wave function has to satisfy boundary condition at specifics points. So, the representation in quasiposition space \cite{pedram2},
\begin{equation}
\label{qp-repx}
\langle x^{ML} | \hat{x} | \psi(t) \rangle = x \psi^{qp} (x,t),
\end{equation}
\begin{equation}
\label{qp-repp}
\langle x^{ML} | \hat{p} | \psi(t) \rangle = -i \hbar \left( 1 - \beta \hbar^{2} \frac{\partial^{2}}{\partial x^{2}} \right) \frac{\partial \psi^{qp} (x,t)}{\partial x},
\end{equation}
to first-order in $\beta$ parameter, is more appropriate\footnote{P. Pedram \cite{pedram2} has proposed a representation in which $\hat{x} = \hat{x}_{o}$ and $\hat{p} = \frac{\tan \left( \sqrt{\beta}\hat{p}_{o} \right)}{\sqrt{\beta}}$, where $\hat{x}_{o}$ and $\hat{p}_{o}$ are ordinary operators of position and momentum, which obey the canonical commutation relation $[\hat{x}_{o},\hat{p}_{o}] = i \hbar$.}. $ |x^{ML}\rangle $ are state vectors of maximal localization which satisfy \cite{Kempf:1994su}
\begin{equation}
\langle x^{ML} | \hat{x} | x^{ML} \rangle = x, \mbox{with} \ x \in \Re,
\end{equation}
\begin{equation}
\left( \Delta x \right)_{| x^{ML} \rangle} = \Delta x_{min} = \hbar \sqrt{\beta},
\end{equation}
and
\begin{equation}
\langle x^{ML} | x^{ML} \rangle = 1.
\end{equation}

The time-independent Schroedinger equation for a non-relativistic particle of mass $m$ in quasiposition space representation takes the form
\begin{equation}
\label{SchrEq}
-\frac{\hbar^{2}}{2m}\frac{d^{2} \varphi^{qp}(x)}{dx^{2}} + \beta \frac{\hbar^{4}}{3m}\frac{d^{4} \varphi^{qp}(x)}{dx^{4}} + V(x)\varphi^{qp}(x) = E \varphi^{qp}(x).
\end{equation}

The above modified Schroedinger equation shows that GUP effects are performed by fourth-order derivative term. This term modifies the probability current as follows\footnote{From now on, we are going to omit the $qp$ superscript of the wave function for sake of simplicity.} \cite{vagenas}
$$
J = - \frac{i \hbar}{2m} \left( \psi^{*}\frac{\partial \psi}{\partial x} - \psi \frac{\partial \psi^{*}}{dx} \right) +
$$
\begin{equation}
\label{probc}
\frac{i \beta \hbar^{3}}{m} \left[ \left( \psi^{*}\frac{\partial^{3} \psi}{\partial x^{3}} - \psi \frac{\partial^{3} \psi^{*}}{\partial x^{3}} \right) +
\left( \frac{\partial^{2} \psi^{*}}{\partial x^{2}} \frac{\partial \psi}{\partial x} - \frac{\partial^{2} \psi}{\partial x^{2}} \frac{\partial \psi^{*}}{\partial x} \right) \right],
\end{equation}
but it does not modify the probability density\footnote{That is because the authors assume that there is no changes in the time-dependent part of the Schroedinger equation.} ,
\begin{equation}
\label{probd}
\rho = | \psi |^{2}.
\end{equation}


\section{Dirac $\delta$-function potential}
\label{Ddfp}

\h In this section, we consider a non-relativistic particle of mass $m$ in the presence of Dirac delta-function potential in a minimal-length scenario. In according to Eq. (\ref{SchrEq}) we have
\begin{equation}
\label{Eqdelta}
-\frac{\hbar^{2}}{2m}\frac{d^{2} \varphi(x)}{dx^{2}} + \beta \frac{\hbar^{4}}{3m}\frac{d^{4} \varphi(x)}{dx^{4}} - V_{0} \delta(x)\varphi(x) = E \varphi(x),
\end{equation}
where $V_{0} > 0$ is a constant.

Integrating Eq. (\ref{Eqdelta}) between $- \epsilon$ and $\epsilon$ (with $\epsilon$ arbitrarily small and positive), and then taking the limit $\epsilon \rightarrow 0$, we obtain

$$
\left[ \frac{d \varphi_{II}(0)}{d x} - \frac{d \varphi_{I}(0)}{d x} \right] - 
$$
\begin{equation}
\label{d3d1}
\frac{2}{3}\beta \hbar^{2} \left[ \frac{d^{3} \varphi_{II}(0)}{d x^{3}} - \frac{d^{3} \varphi_{I}(0)}{d x^{3}} \right] + \frac{2mV_{0}}{\hbar^{2}} \varphi(0) = 0,
\end{equation}
where $\varphi_{I}(x)$ and $\varphi_{II}(x)$ are the solutions of Eq. (\ref{Eqdelta}) for $x < 0$ and $x > 0$, respectively.

Since the third derivative of $\varphi(x)$ at $x =0$ has a finite discontinuity (that is to say, a jump by a finite amount), we require that the second and first derivatives are continuous at $x = 0$. Consequently, Eq. (\ref{d3d1}) turns into \cite{robinett,robinett1}
\begin{equation}
\label{d3d}
\frac{\beta}{3} \left[ \frac{d^{3} \varphi_{II}(0)}{d x^{3}} - \frac{d^{3} \varphi_{I}(0)}{d x^{3}} \right] = \frac{mV_{0}}{\hbar^{4}} \varphi(0).
\end{equation}

As it is well-known, taking into account the sign of the energy two case can then arise: (i) bound states when $E < 0$ and (ii) scattering states when $E > 0$.

\subsection{Bound states}
\label{BS}

\h In this case, the general solution of Eq.(\ref{SchrEq}) is given by
\begin{equation}
\label{wf1bs}
\varphi_{I,II}(x) = A_{I,II} e^{kx} + B_{I,II} e^{-kx} + C_{I,II} e^{k_{\beta}x} + D_{I,II} e^{-k_{\beta}x},
\end{equation}
where, to first order in $\beta$,
\begin{equation}
\label{k}
k := k_{0} \left( 1 + \frac{1}{3} \beta \hbar^{2} k_{0}^{2} \right), 
\end{equation}
\begin{equation}
\label{k_beta}
k_{\beta} := \sqrt{\frac{3}{2 \hbar^{2} \beta}} \left( 1 - \frac{1}{3} \beta \hbar^{2} k_{0}^{2} \right) 
\end{equation}
and
\begin{equation}
\label{k_zero}
k_{0} := \sqrt{\frac{2 m |E|}{\hbar^{2}}}. 
\end{equation}

The coefficients can be found, except by one normalization constant, requiring that solutions remain finite when $x \rightarrow \pm \infty$ and the continuity of solution and of its first and second derivatives at $x = 0$. We come to the result
\begin{equation}
\label{bss}
	\left \{
	\begin{array}{ll}
	\varphi_{I}(x) = A e^{kx} - \frac{k}{k_{\beta}} A e^{k_{\beta}x}, & x < 0 \\ [13pt]
	\varphi_{II}(x) = A e^{-kx} - \frac{k}{k_{\beta}} A e^{-k_{\beta}x}, & x > 0, \\
	\end{array}
	\right.
\end{equation}
where $A$ is the normalization constant.

From Eq. (\ref{d3d}) we can find the bound state energy up to order $\beta$ as 
\begin{equation}
\label{bsenergy}
E = - \frac{m V_{0}^{2}}{2 \hbar^{2}} + \sqrt{ \frac{2 \beta}{3}}\frac{m^{2} V_{0}^{3}}{\hbar^{3}} - 2 \beta \frac{ m^{3} V_{0}^{4}}{\hbar^{4}},
\end{equation}
which is in agreement with N. Ferkous's result \cite{Ferkous}. It is interesting to note that the first correction brought about by the introduction of a minimal-length scenario is ${\sl O}(\sqrt{\beta})$.

\begin{figure}
\centering
\includegraphics[scale=1.0]{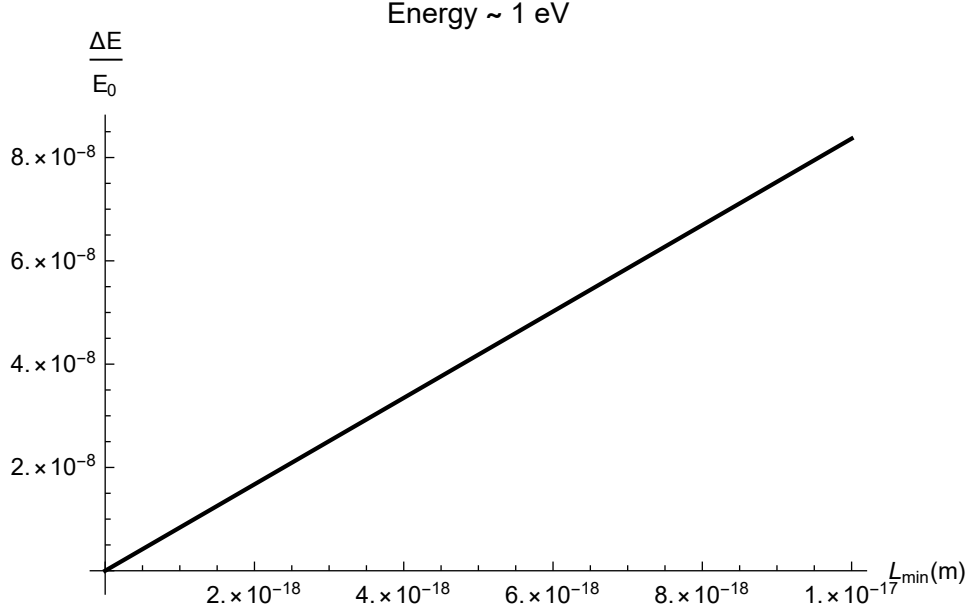} 
\caption{Bound state: $\frac{\Delta E}{E_{0}}$ as function of $L_{min}$ in units of meter, for $E_{0} \approx 1$ eV.}
\label{figligml}
\end{figure}

\begin{figure}
\centering
\includegraphics[scale=1.0]{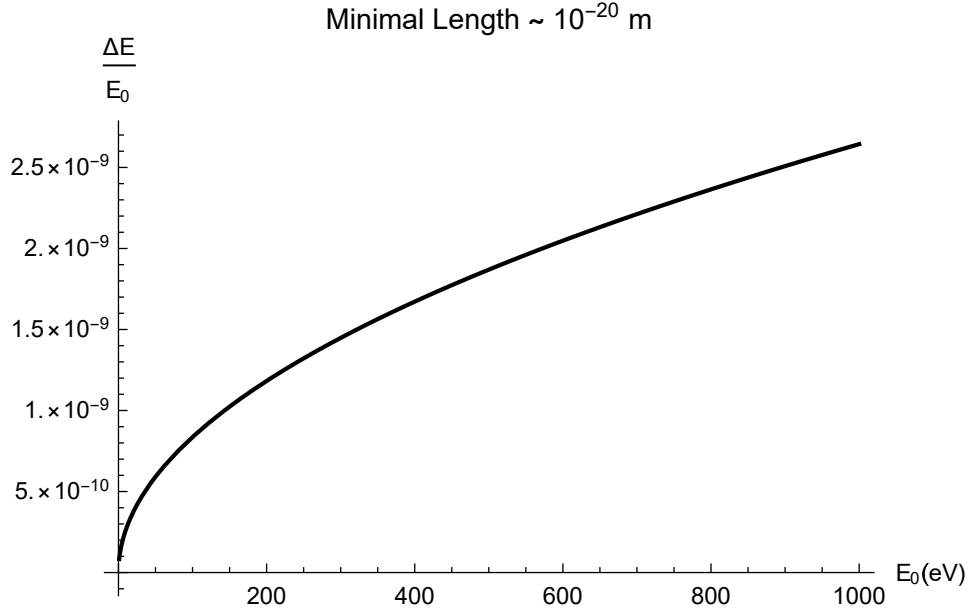} 
\caption{Bound state:$\frac{\Delta E}{E_{0}}$ as function of $E_{0}$ in units of eV,  for $L_{min} = 10^{-20}$ m.}
\label{figligener}
\end{figure}

For an electron, the relative difference between the bound state energy arising from the introduction of a minimal length and the absolute value of the ordinary energy of the bound state is showed as a function of the minimal length for the energy about 1 eV in Fig. \ref{figligml} and as a function of $E_{0}$ (1 eV $\leq E_{0} \leq$ 1 keV ) for $L_{min} = 10^{-20}$ m in Fig. \ref{figligener}. In Fig. \ref{figligml} we choose the $10^{-17}$ m upper value for the minimal length because it is in accordance with that commonly found in the literature \cite{brau,quesne,antonacci} and it is consistent with the one at the electroweak scale \cite{tawfik2,das1,das2}. For the Planck's length, $\frac{\Delta E}{E_{0}} \approx 8.4 \times 10^{-26}$, unfortunately a virtually unmeasurable effect quantum gravity using current technology.

\subsection{Scattering states}
\label{SS}

\h In this case, the general solution of Eq.(\ref{SchrEq}) is given by
\begin{equation}
\label{wf1ss}
\varphi_{I,II}(x) = A_{I,II} e^{ikx} + B_{I,II} e^{-ikx} + C_{I,II} e^{k^{\prime}_{\beta}x} + D_{I,II} e^{-k^{\prime}_{\beta}x},
\end{equation}
where
\begin{equation}
\label{k_beta_linha}
k^{\prime}_{\beta} := \sqrt{\frac{3}{2 \hbar^{2} \beta}} \left( 1 + \frac{1}{3} \beta \hbar^{2} k_{0}^{2} \right).
\end{equation}

Now we demand there is not reflected wave function for $x > 0$, consequently $ B_{II} = 0$. From requirement that solutions remain finite when $x \rightarrow \pm \infty$ we have $D_{I} = 0$ and $C_{II} = 0$. In this case, the continuity of solution and of its first and second derivatives at $x = 0$ are not enough to find the coefficients. It is also necessary to use the discontinuity of the third derivative at $x = 0$, Eq. (\ref{d3d}). After some algebra, we have
\begin{equation}
\label{sss}
	\left \{
	\begin{array}{ll}
	\varphi_{I}(x) = A e^{ikx} + \frac{ik^{\prime}_{\beta}}{k} \frac{A}{b}e^{-ikx} - \frac{A}{b}e^{k^{\prime}_{\beta}x}, & x < 0 \\ [13pt]
	\varphi_{II}(x) = \frac{aA}{b} e^{-ikx} - \frac{A}{b}e^{-k^{\prime}_{\beta}x}, &  x > 0, \\
	\end{array}
	\right.
\end{equation}
where
\begin{equation}
\label{a}
a := 1 +\frac{2 \beta \hbar^{4} k^{\prime}_{\beta}}{3 m V_{0}} \left(k^{\prime 2}_{\beta} + k^{2} \right),
\end{equation}
\begin{equation}
\label{b}
b := a - i \frac{k^{\prime}_{\beta}}{k},
\end{equation}
and $A$ is a normalization constant.

Consequently, the reflection and transmission coefficients are given by
\begin{equation}
\label{reflectionc}
R = \left( \frac{k^{\prime}_{\beta}}{k} \right)^{2} 
\frac{1}{ \left[  1 +\frac{2 \beta \hbar^{4} k^{\prime}_{\beta}}{3 m V_{0}} \left(k^{\prime 2}_{\beta} + k^{2} \right) \right]^{2} + \left( \frac{k^{\prime}_{\beta}}{k} \right)^{2}}
\end{equation}
and
\begin{equation}
\label{transc}
T = \left( \frac{k^{\prime}_{\beta}}{k} \right)^{2} \frac{\left[  1 +\frac{2 \beta \hbar^{4} k^{\prime}_{\beta}}{3 m V_{0}} \left( k^{\prime 2}_{\beta} + k^{2} \right) \right]^{2}}{ \left[  1 +\frac{2 \beta \hbar^{4} k^{\prime}_{\beta}}{3 m V_{0}} \left(k^{\prime 2}_{\beta} + k^{2} \right) \right]^{2} + \left( \frac{k^{\prime}_{\beta}}{k} \right)^{2}}.
\end{equation}
Note that $R + T = 1$, as must be.

It is instructive  to write the reflection and transmission coefficients up to first corrections. Then,
\begin{equation}
\label{reflectionc1}
R = \left( 1 + \frac{2\hbar^{2} |E|}{m V_{0}^{2}} \right)^{-1} \left[ 1 - \sqrt{\frac{2 \beta}{3}} \frac{2mV_{0}}{\hbar}\left( 1 + \frac{m V_{0}^{2}}{2\hbar^{2} |E|} \right)^{-1} \right]
\end{equation}
and
\begin{equation}
\label{transc1}
T = \left( 1 + \frac{m V_{0}^{2}}{2\hbar^{2} |E|} \right)^{-1} \left[ 1 + \sqrt{\frac{2 \beta}{3}} \frac{m^{2}V_{0}^{3}}{\hbar^{3} |E|}\left( 1 + \frac{m V_{0}^{2}}{2\hbar^{2} |E|} \right)^{-1} \right].
\end{equation}

Above results show that the reflection and the transmission coefficients are no longer the same in the cases of a delta-function well ($V_{0} > 0$) and a delta-function barrier ($V_{0} < 0$). Therefore the presence of a minimal length decreases the chances of tunneling.

It is also interesting to note that the first correction brought about by the introduction of a minimal-length scenario is ${\sl O}(\sqrt{\beta})$ in the same way as in the bound state energy.

\begin{figure}
\centering
\includegraphics[scale=1.0]{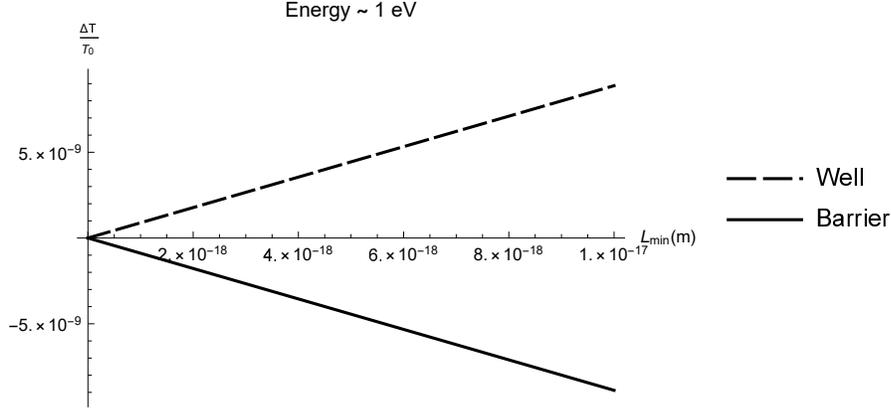} 
\caption{Scattering state: $\frac{\Delta T}{T_{0}}$ as function of $L_{min}$ in units of meter, for $E_{0} \approx 1$ eV.}
\label{tunev}
\end{figure}

\begin{figure}
\centering
\includegraphics[scale=1.0]{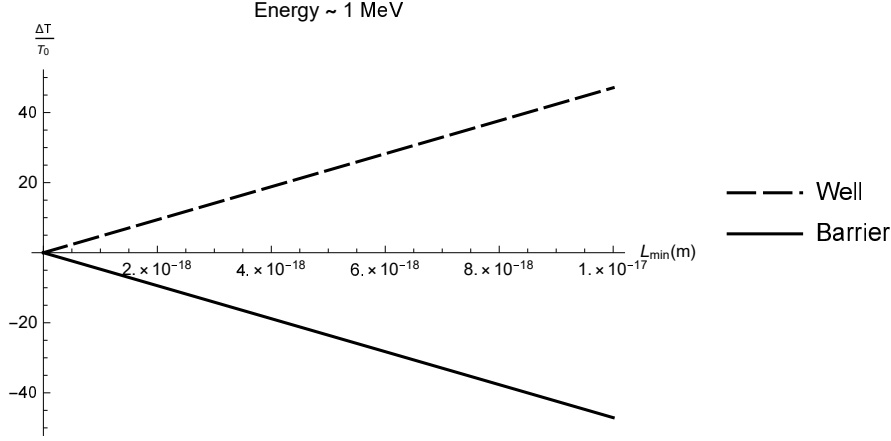} 
\caption{Scattering state: $\frac{\Delta T}{T_{0}}$ as function of $L_{min}$ in units of meter, for $E_{0} \approx 1$ MeV.}
\label{tunmev}
\end{figure}

Fig. \ref{tunev} and Fig. \ref{tunmev} show the relative difference between the transmission coefficient arising from the introduction of a minimal length and $T_{0}$ (ordinary transmission coefficient) for the cases of a Dirac delta well (dashed line) and of a Dirac delta barrier (continuous line). Fig. \ref{tunev} is for electrons scattering of energy about 1 eV and $V_{0} = 2$ eV\AA. For the Planck's length, $\frac{\Delta T}{T_{0}} \approx 8.9 \times 10^{-27}$, again a virtually unmeasurable effect. Fig. \ref{tunmev} is for protons scattering of energy about 1 MeV and $V_{0} = 3 \times 10^{-2}$ MeV\AA. For the Planck's length, $\frac{\Delta T}{T_{0}} \approx 4.7 \times 10^{-17}$. Note that $L_{min} \sim 10^{-17}$ m results in significant effects, which may be an indication that $L_{min}$ is far from the electroweak scale.

\subsection{Remarks}
\label{remarks}
\begin{enumerate}
\item{It is easy to see that in the limit $\beta \rightarrow 0$ we recover the results known for Dirac $\delta$-function potential in ordinary quantum mechanics.}

\item{A more detailed analysis shows that $k_{\beta}$ and $k^{\prime}_{\beta}$ do not vanish even if $m = 0$. Therefore, $e^{-k_{\beta}|x|}$ and $e^{-k^{\prime}_{\beta}|x|}$ solutions still persist since $ e^{-k_{\beta}|x|}$, $e^{-k^{\prime}_{\beta}|x|} \rightarrow e^{-\sqrt{\frac{3}{2 \hbar \beta}}|x|}$ when $m = 0$. Consequently, this leads us to presume that such solutions are ``background solutions'' caused by introduction of an effective description of the effects of quantum gravity. However, since their coefficients in Eqs. (\ref{bss}) and (\ref{sss}) vanish when $m = 0$, they are not present in the bound state and the scattering states solutions.}

\item{It is important to point out that now the first derivative at $x = 0$ is no longer discontinuous. However, in the limit $\beta \rightarrow 0$ the discontinuity at $x = 0$ is recovered. Moreover, if the term of ${\sl O}(\beta^2)$ is considered in the Schroedinger equation the third derivative will turn into continuous at $x = 0$, and so on.}

\item{$e^{-k_{\beta}|x|}$ and $e^{-k^{\prime}_{\beta}|x|}$ solutions are only significant for very small values of $x$, that is, high energy. Thus we could assume that they lie far outside validity range at which the Schroedinger equation may consistently work and throw them away. However, that is a naive assumption, because they lead to the emergence of traces of quantum gravity in low energy physics, as the previous results show. Note that they provide the continuity of first and second derivatives at $x = 0$.}

\item{It is known, at least since the Frost's work of 1954 \cite{frost}, that the ground state energy of the hydrogen atom (1s state) is identical to the bound state energy of a Dirac $\delta$-function potential in 1-dim when $V_{0}$ is replaced by the fine structure constant, $\alpha$. Thus, assuming that this identiy is kept in a minimal-length scenario\footnote{ Since the symmetry of the 1s state must remain the same in both cases.}, the result (\ref{bsenergy}) predicts a leading correction for ground state energy of the hydrogen atom of ${\sl O}( \sqrt{\beta})$, whereas the result commonly found in the literature using perturbative methods is of ${\sl O}(\beta)$ \cite{brau,yao,benczik,Stetsko,antonacci}.

It is importante to add that using a non-pertubative approach T. V. Fityo, I. O. Vakarchuk and V. M. Tkachuk \cite{fityo} and  D. Bouaziz \& N. Ferkous \cite{bouaziz} have also found a first correction of  ${\sl O}( \sqrt{\beta})$.

Now, we can make a rough estimate of an upper bound for the minimal-length value comparing our result with experimental data \cite{antonacci}. Using data obtained in reference \cite{parthey}, in which the accuracy of about $4,2 \times 10^{-14}$ eV has been obtained, we find that $\Delta x_{min} \le 10^{-25}$ m. Hence, in the case of the protons scattering from the previous subsection, we find $\frac{\Delta T}{T_{0}} \sim 10^{-17}$ for $L_{min} \sim 10^{-25}$ m, which is a more representative result.
}

\end{enumerate}


\section{Conclusion}
\label{Concl}

\h In this work, we solve, in quasiposition space, the Schroedinger equation for a Dirac $\delta$-function potential. Our result for the bound sate energy is in agreement with that calculated by Ferkous in momentum space. Moreover, we find that leading correction for the reflection and transmission coefficients of the scattering states, the bound state energy and ground state of the hydrogen atom are of order of the minimal length, ${\sl O}(\sqrt{\beta})$. We also show that in the presence of a minimal length the coefficients of reflection and transmission for the Dirac delta-function well and the Dirac delta-function barrier are no longer the same. There is a decrease in the chances of tunneling.

Although different physical systems can be modeled by a Dirac $\delta$-function potential, we have to ask ourselves of the validity of the results, since the Dirac $\delta$-function potential is already an approximation to an actual physical system. That is, are the minimal-length effects smaller than the ones due to the modeling by the Dirac $\delta$-function potential? Probably the answer is yes, though it is difficult to insure. What we can claim is the estimates of a upper bound for the minimal-length value are acceptable, in the sense that even though the corrections for a more realistic potential can be greater than ones due to the minimal-length effects, that only leads to upper bound values even smaller. However that is not very different from others systems we have studied in a minimal-length scenario.



\section*{Acknowledgements}
We would like to thank FAPES, CAPES and CNPq (Brazil) for financial support.
\\

\end{document}